\documentclass{elsart}
\usepackage{graphics}
\usepackage{graphicx}
\usepackage{epsfig}
\usepackage{amssymb}

\date{}
\begin{document}
\begin{frontmatter}

\title{Magnetism, FeS colloids, and Origins of Life}
\emph{Dedicated to the memory of Professor Alladi Ramakrishnan}

\author{Gargi Mitra--Delmotte\corauthref{cor1}}
\corauth[cor1]{Corresponding author; Present Address: 39 Cite de
l'Ocean, Montgaillard, St. Denis 97400, REUNION. Telephone and Fax
no. 00-262-262307972; Email: gargijj@orange.fr} and
\author{A.N.Mitra\corauthref{cor2}} \corauth[cor2]{Formerly INSA Einstein Professor, Department of
Physics, Delhi University; address : 244 Tagore Park, Delhi-110009.
INDIA; Email: ganmitra@nde.vsnl.net.in}

\begin{abstract}
A number of features of living systems: reversible interactions and
weak bonds underlying motor-dynamics; gel-sol transitions; cellular
connected fractal organization; asymmetry in interactions and
organization; quantum coherent phenomena; to name some, can have a
natural accounting via $physical$ interactions, which we therefore
seek to incorporate by expanding the horizons of `chemistry-only'
approaches to the origins of life. It is suggested that the magnetic
'face' of the minerals from the inorganic world, recognized to have
played a pivotal role in initiating Life, may throw light on some of
these issues. A magnetic environment in the form of rocks in the
Hadean Ocean could have enabled the accretion and therefore an
ordered confinement of super-paramagnetic colloids within a
structured phase. A moderate H-field can help magnetic
nano-particles to not only overcome thermal fluctuations but also
harness them. Such controlled dynamics brings in the possibility of
accessing quantum effects, which together with frustrations in
magnetic ordering and hysteresis (a natural mechanism for a
primitive memory) could throw light on the birth of biological
information which, as Abel argues, requires a combination of order
and complexity. This scenario gains strength from observations of
scale-free framboidal forms of the greigite mineral, with a magnetic
basis of assembly. And greigite's metabolic potential plays a key
role in the mound scenario of Russell and coworkers-an expansion of
which is suggested for including magnetism.
\end{abstract}
\end{frontmatter}

% 1.0

\section{Introduction}
Life's hierarchical control structure is a sequence of constraints,
each limiting the scope of the preceding level for step-wise
harnessing of the physico-chemical laws governing its lowest rung.
But the limiting 'boundary conditions' are themselves extraneous;
they cannot be formally derived from these laws. Further, the
higher-level operating principles depend on, but are $not$ reducible
to, those of the lower ones (Polyani 1968). Next, the origins of
purpose permeating across biology (Kant 1790), as well as
information associated with function, are among the most fundamental
of questions in biology (K\"uppers 1990). Indeed, the
structure-function relationship where rate-dependent equations
representing measurement associated with bio-structures are linked
to rate-independent constraints associated with bio-information, is
viewed as an epistemological complementarity (Pattee 1979).
According to Pattee, "epistemic operations like observation,
detection, recognition, measurement, and control as the essential
type of function" demarcate living from non-living organizations.
The chances of an organism's survival are crucially dependent on its
ability to improve its control strategies that in turn depend on its
recognition of environmental patterns. Hence, "To qualify as a
measuring device it must have a function, and the most primitive
concept of function implies improving fitness of an organism".
Pattee's famous "semantic closure principle", places a heavy
responsibility on the $observer$ who should at minimum be an
$organization$ that can construct the measuring device and use the
results of measurement for its very survival (Pattee 1996). This
scenario seems to be a far cry from the objective
(observer-independent) physical laws characterized by Universality
and Invariance Principles. And it is indeed a tall order to explain
from these `classical' premises the emergence of subjective
(observer-dependent) biological infrastructure making measurements
for survival. But Pattee recognizes that unlike classical theory,
Quantum theory is $not$ constrained by observer - independence and
promptly invokes Wheeler to make his point:  "No elementary quantum
phenomenon is a phenomenon until it is a recorded phenomenon (i.e.,
the results of a measurement)". Indeed, the puzzle is really about
how the `Cybernetic cut' (Abel 2008) could have been crossed using
mere physico-dynamics, leading to the emergence of a non-physical
(not governed by chance or necessity) mind from physicality that
established controls over the same. We further ask if this mystery
could somehow be related to the idea of life having originated in an
inorganic world-an idea which has met considerable acceptance. The
compelling link to iron-sulphide (FeS) clusters in early evolved
enzymes (and across species in a range of crucial roles, e.g.
catalytic, electron transfer, structural), with exhalates on the
Hadean ocean floor, is based on the close resemblance of these
clusters with greigite (Fe$_{5}$NiS$_{8}$) (Russell et al 1994;
Huber and W\"achtersh\"auser 1997). Not only are these clusters seen
as playing a key role in the origins of metabolism, where
geochemical gradients were harnessed, but also, for long mineral
crystal surfaces have, and continue to be seen as scaffolds thanks
to their chemical-information storing/transferring potential,
leading to the other-replicating-- wing of Life (Bernal 1949;
Goldschmidt 1952; Degens et al 1970; Cairns-Smith 1982; Arrhenius
2003; Ferris 2005). But in these approaches, a number of features:
reversible interactions, weak bonds, gel-sol transitions, cellular
connected fractal organization, asymmetry in interactions and
organization, to name some, and which are difficult to address using
chemical interactions alone, are seen as later arrivals, i.e. upon
achievement of complexity in the pre-biotic 'soups'. Here again, the
path, as to how complexity could have been entrained to lead to
Life-like features of today, remains far from being understood.
Then, in addition to chemistry, could physical properties of
inorganic matter have also acted as a scaffold for onward
transmission of several common $physical$ features (see below)
typical of living systems? To that end, we note that dynamically
ordered forms of matter, like framboids, regardless of chemical
structure are the result of physical forces, including magnetism
(see Sect. 4).

Now, magnetism has myriad manifestations at different scales
--quantum to cosmological (Skomski 2008). (The repeated appearance
of fractal themes is compelling -- from magnetic critical phenomena
to finer length scales where quasiparticle behaviour in a magnetic
field can be explained by fractional quantum numbers (Jain 2007;
Goerbig et al. 2004); Farey series elements, $F_n$ ; Hausdorf
dimension $h$ (da Cruz 2005)). And, there are ubiquitous magnetic
influences across kingdoms : navigation sensing in bacteria, algae,
protists, bees, ants, fishes, dolphins, turtles and birds
(Kirschvink and Gould 1981; Winkelhofer 2005); field effects on
growth patterns, differentiation, orientation of plants and fungi
(Galland and Pazur 2005); ferromagnetic elements in tissues
(Kirschvink et al. 1992), etc. (In fact, magnetite ($Fe_3O_4$, a
magnetic mineral) - biomineralization, the most ancient
matrix-mediated system, is thought to have served as an ancestral
template for exaptation (Kirschvink and Hagadorn 2000)). Indeed, new
inputs of quantum events underlying biophenomena like
magnetoreception (Kominis 2008) reveal the importance of magnetism
in biological systems of today. Most importantly, its vital role in
{\it the science of information technology} persuades us to turn to
this enveloping science for any mechanisms beyond the limits of
physico-chemical principles that could have helped bridge the gap
from inanimate matter to life.

\noindent In this mini survey
\begin{enumerate}
\item We give a brief summary of the relevance of quantum searches in biology and therefore to the origin-of-life problem (Sect. 2.1). We briefly review spin and magnetic models offering insights into the emergence of life, leading up to our proposal (Sects.2.2-4).
\\
\item We survey various biophenomena with analogies to magnetic ones in general as well as topological similarities with our magnetism-based proposal in particular (Sects.3.1-9), and ask if magnetism could have helped to pave the way for a take-off from non-life to life.
\\
\item We briefly review framboids, where conflicting physical forces usher in dynamic order. Here, the mineral greigite's magnetic properties underlie its framboid-forming capacity (Sect.4).
\\
\item We outline the mound scenario of Russell and coworkers, with rich metabolism potential, where greigite forms in a colloidal environment. A possible scenario for a magnetic reproducer is drawn (Sect.5). \\
\end{enumerate}

% 2.0
\section{Quantum searches and the origins of life}
A brief introduction on quantum searches in biology is followed by
their implications in the origin of life. A possible physical system
enhancing the propensity of such searches is then suggested. 2.1
\subsection{Quantum searches and biology}
Outstanding biological-search examples can be seen in biological
evolution itself, with divergences symbolized by tree nodes; the
clonal Darwinian-like phase in the adaptive immune system; brain
connections and protein folding. The efficiency of quantum searches
over classical ones has prompted the idea that they could have been
used by Nature who usually is found to take the cleverest among
available options, as illustrated by certain Extremum Principles of
Classical physics (Hamilton, Fermat, Maupertius). For instance, in a
database of dimension $d$, a quantum search gives a square root
speed-up over its classical counterpart -- also valid for the
respective nested versions (Cerf et al. 2000). In a typical
scenario, challenges interrupting the networking phase are seen as
forcing the biosystem to seek help from a co-existing quantum
domain, e.g., a search prompted by a 'crisis' in the form of a
depleted nutrient could lead the adaptive system to a new pathway
for succour. Now, quantum coherence in the set of elements on the
affected front could help skirt frustrations in local minima as can
happen in a classical search. This access to the wave-property
enables a superposition of states and allows a `holistic' decision.
Thus in the face of crises, halted networked interactions in a
subsystem would prompt the formation of a `quantum decision front'.
This would be constantly checked or 'measured' by the rest of the
system. A fruitful interaction with one chosen path would mean a
simultaneous collapse of the quantum superposition of alternative
paths (McFadden and Al-Khalili 1999).

Today, clear signatures of quantum processing in biology are coming
in ((Engel et al. 2007), aided by femtosecond laser-based 2D
spectroscopy and coherent control approaches, showing how phase
relationships in nano-structures modulate the course of
bio-reactions (Nagya et al 2006). As to decoherence evading
mechanisms, the role of a gel-state; quasicrystalline order; (Jibu
et al 1994; Hagan et al. 2002); are amongst proposed
order-maintaining mechanisms in a wet environment, while 'screening
effect', or 'cocooning' structural mechanisms are seen as providing
insulation against interactions with the environment [Patel (2001);
Davies (2003, 2004)] (see also Sect. 3.9). Indeed, it seems that
Nature has quietly been using these strategies all along, i.e.
leading to creation of biological language itself, as the
Grover-Patel search numbers match those used by Nature ! Using
Grover's quantum search method for a marked item in an unsorted
database, Patel (2001) hit upon the base-pairing logic of nucleic
acids in transcription and translation as an excellent quantum
search algorithm -- a directed walk through a superposition of all
possibilities - resulting in a 2-fold increase in sampling efficacy
over its classical counterpart (which at best permits a random
walk). Prompted by these insights, Al-Khalili and McFadden (2008)
point out that a quantum search would have been far more efficient
than a random one for picking out the self-replicator from the
primordial soup comprising a dynamic combinatorial library of
compounds linked together, say by reversible reactions. But what
plausible ingredients could have facilitated such a quantum assisted
leap?

% 2.2

\subsection{Spin and magnetic systems for the origin-of-life}
Hypothesizing a quantum-mediated process for the transition from
non-life to life, Davies (2008) proposes that information could have
its origins in quantum objects such as spins, whose orientations
offer a natural discretization mechanism of genetic information, and
which in turn may have been embodied by physical structures in some
natural system.  Although this would initially be copying bits (no
associated phase information so initially no issues of decoherence
evasion) the possibility of coherence in this inherently quantum
system endows it with potential for conducting a quantum search for
the quantum replicator. Furthermore, he points out that in this
envisioned scenario, the collapse of the quantum superposition of
states of living and non-living ones to the low probability state of
"life", cannot be due to the quantum system's own doing. Instead it
must have been the result of an environmental interaction, serving
as a measuring device, thus implying a key role for the environment
(c.f. Zurek 2003). Again, an origin-of-life model based on
spin-ordering (a variant of the Ising spin glass) was proposed by
Anderson (1983), that was albeit prompted from another angle -- the
correspondence between the complexity due to the impact of
frustration in magnetically disordered systems and bio-processes,
such as protein-folding (Hollander and Toninelli 2004; Stein 1996)
(see Sect. 3.1). Then again,  Breivik (2001) demonstrated that
self-ordering of ferromagnetic objects ($\sim$ 3mm) with
reproduction of magnetic templates could be manipulated via dynamic
interaction with environmental temperature fluctuations, thereby
significantly also connecting information encoded in nucleic acids
with non-chemically linked aperiodic polymers. This is because a
magnetically packed array is naturally aperiodic (see Sect. 3.8),
hence \textit{satisfying Schroedinger's (1944) vision of aperiodic
surfaces as efficient information-holders}, in contrast to a
periodic crystal lattice with strongly correlated elements. This
magnetic mechanism for propagating information, also agrees with
Dyson's (1999) suggestion that `physical reproduction' preceded
chemical replication in the origins of life, the latter being
identified with a specific chemical copying process. And
interestingly, his use of a magnetic analogy for states, obeying the
Boltzmann probability distribution, gels with the kinetic aspects of
biological reactions (Pross 2005).

All this compels us to ask if magnetism could have empowered the
initial conditions for traversing the bridge dividing life from
non-life, by providing simultaneously a scaffold for interactions
and connections, where physical representations would allow for
higher level abstractions, not of the isolated system but rather in
the context of its penetrating environment playing an active role in
its decision-making. We have recently, proposed (Mitra-Delmotte and
Mitra 2007; 2009) that an external field in the form of magnetic
rocks could have enabled accretion of newly forming, magnetic
nano-particles on the Hadean Ocean floor, since field-induced
aggregates have been observed in magnetic fluids showing deviations
from ideal behaviour.

%  2.3
\subsection{Ferrofluids; field-induced structures}
Ferrofluids are colloidal single-domain magnetic nanoparticles
($\sim$ 10nm) in non-magnetic liquids that can be controlled by
moderate H-fields ($\sim$ tens of milliTesla) (Odenbach 2004). The
relevance of these dispersions to natural locations has been
considered only rarely, for e.g. see Wilkins and Barnes (1997),
perhaps due to their synthetic origins; nevertheless their amazing
properties lead to myriad applications, including ratchet behaviour
(Engel et al 2003). Dilute dispersions display ideal single-phase
behaviour due to prohibited (chemical) inter-particle contacts,
thanks to synthetic coatings. On the other hand, in the present
context we look at the interactions between the magnetic particles
although the carrier remains in the liquid state. Such deviations
from ideal magnetization behaviour can show up on increasing
particle concentrations that can be understood in terms of
H-field-induced inter-particle interactions leading to internal
structure formation (Rosenweig 1997; Chantrell et al 1982) and
manifesting in {\it dense phases} -a milder phase transition than to
the solid-crystalline one. The structure of hydrated, heterogenous
aggregates would depend on factors like the strength of the applied
field, the nature of the ferrofluid, etc. (Odenbach 2004; Zubarev et
al 2005; Zubarev and Iskakova 2004). Li et al (2007) have pointed
out the dissipative nature of the field-induced aggregates (Taketomi
et al (1991)) that break up in response to thermal effects upon
removal of field. In their gas-like compression model, the total
magnetic energy of ferrofluids obtained from an applied field: $W_T
= W_M + W_S $; where $ W_M = \mu_{0} M H V $ and $W_S = -T \Delta S
$ are the magnetized and the structurized energies, respectively,
$V$ is the volume of the ferrofluid sample and $\Delta S $ is the
entropic change due to the microstructure transition of the
ferrofluid. An assumed equivalence of $W_T$ (zero interparticle
interactions), with the Langevin magnetized energy $W_L = \mu_{0} M
H V $ necessitates to a correction in the magnetization, in terms of
the entropy change. Hence, these colloidal systems are well equipped
to analyze the interplay between competing factors -dipolar
interactions, thermal motion, screening effects, etc. leading to the
emergence of magnetically structured phases (Pastor-Satorras and
Rubi 2000).

% 2.4
\subsection{Structured magnetic phases; life-like dynamics}
On analogous lines to ferrofluids, magnetic rocks providing a
surface field strength $\sim$ tens of milli-Tesla would have turned
any newly forming magnetic particle suspension into tiny magnets,
leading to the emergence of magnetically structured phases (MSPs).
We come to a suggested scenario in Sect.5. Here the magnetic entropy
property of super-paramagnetic particles offer a ready basis for
interchange with the Brownian hits from the surroundings for
harnessing this energy, analogous to complex biological soft matter,
while the external magnetic environment plays a key role in
controlling their dynamics. Further, we suggested (Mitra-Delmotte
and Mitra 2009) that the presence of charge on particles would
permit only the tiny sized particles (carrying one/two units of
charge) to diffuse through layers of the magnetically accreted
charged layers in response to a non-equilibrium source-- a gentle
gradient of flux lines (assuming a non-homogeneous H-field from
rocks). Non-equilibrium energy driven diffusion of tiny particles
(ligand-carrying or otherwise) through the magnetically ordered
phase in a close-to-equilibrium manner shows the possibility of
controlled dynamics in a confined system.

The connections between field-induced structures of magnetic
nano-particles and bio-phenomena bring out their ramifications for
fluctuation-generated order from dissipative structures as envisaged
decades ago (Nicolis and Prigogine 1977). Note that a magnetic
environment exerts control on spin states and hence on
spin-selective chemical reactions (see Buchachenko 2000). The
possibility of yet another magnetic control is via magnetically
sensitive reactions whose rates are sensitive to orientations of
reactants (Weaver et al 2000). Separation of complex mixtures
forming at the origins of life would have also been facilitated by
magnetic mechanisms, acting in an orthogonal non-interfering manner.

% 3.0
\section{'The importance of being magnetic'}
We now look at some general features of biological systems with
similarity to magnetic phenomena, also comparing dynamics in biology
vis-a-vis our proposal of a nano-scale assembly controllable by a
magnetic environment.

% 3.1
\subsection{Confinement, connectivity, frustration-complexity}
Self-ordering phenomena (Nicolis and Prigogine 1977) show how
spontaneous order can emerge from inanimate matter, leading to
connected components (confined). But the high algorithmic
compressibility of order and patterns that can be explained in terms
of physical laws would simultaneously make it difficult to generate
the complexity (high information carrying capacity) underlying
biology (Abel 2009). In Shannon's terminology, the information
carrying capacity of a 1D-string is at its maximum when there are no
correlations between its components, i.e. when it is a random
sequence. A combination of the two-order and unpredictability -
might be a better way to understand this paradox of biological
complexity (Abel 2009). Now, frustrations in magnetically connected
systems are well known in literature (see also Sects. 2.2, 4.5).
Their presence, naturally introduce the element of uncertainty in
the midst of long-range correlations. We therefore suggest that a
confined system due to magnetic connections, as in our proposal, has
the combination for addressing such complexity in the origins of
life.

% 3.2
\subsection{Nested hierarchy, cooperative dynamics}
Biological structures appear as nested organizations based on
coherent feedback through a lattice of interacting, spatially
oriented units; self and non-self interactions underlie their
cooperative dynamics [Ling (2001)]. And as noted by Min et al (2008)
the characteristics of dynamically self-assembled nano-structures
with bottom-up complexity, formed by dissipating energy, depend on
the constituent particle size, shape, hardness, composition, apart
from their sensitivity to (control by) external fields; this
approach was used in generating systems with hierarchial complexity
via an interplay of magnetic and hydrodynamic interactions
(Grzybowski and Campbell 2004) (see also Sect.4). In this connection
recall some facets of magnetism in common with those of
self-organizing systems: emergence of global order from local
interactions, organizational closure, hierarchy, downward causation,
distributed control underlying robustness, bifurcations via boundary
conditions, non-linearity due to feedback, etc. (Heylighen 2001).
Their relevance can be gauged from the insights of Bak and Chen
(1991):  long-range spatio-temporal correlations (via a
non-dimensional scale factor) are the hallmark of self-similarity,
manifest as self-organized criticality in natural dynamical systems.
Again, Selvam (1998) proposed a coherence preservation mechanism via
self-similar structures with quasicrystalline order as iterative
principles -- the main tools for handling non-linear dynamics of
perturbations for evolving nested order that connect the microscopic
and macroscopic realms with scale-free structures arising out of
deterministic chaos. This brings us to Tagore's couplet:

\begin{center}
\begin{minipage}{13cm}
{\small {\em
\begin{quote}
Amra shobai raja amader ei rajar rajottey, noiley moder rajar shoney
milbey ki shottey  -- Tagore
\end{quote}
} (We are all kings in our King's kingdom, else how do we get along
with Him.) }
\end{minipage}
\end{center}

% 3.3
\subsection{Polar cell-organization and structures}
On higher scales, the directionality of biochemical processes gets
derived from the asymmetric structure of biomolecules and their
association into consequently polarized assemblies with increasing
complexity [Harold (2005)]. The cytoskeleton, at least in
eukaryotes, is organized via transmitted internal or external
spatial cues, reflecting the polar organization of the cell [Drubin
(2000)]. We also note that some fundamental biological structures
form from asymmetric monomers. For instance, the directionality of
nucleic acid polymers stems from the asymmetry of template-based
aligning monomers. The cytoskeletal family of proteins provides
another outstanding example. The past two decades revealed how
analogous functions are carried out by bacterial homologues of
eukaryotic cytoskeletal proteins. Actually, the highly conserved
FtsZ, barring a few exceptions, is found across all eubacteria and
archaea. Despite its low sequence identity to tubulin, its
eukaryotic homologue, the two proteins not only share the same fold
but follow similar self-assembly patterns, forming protofilaments.
The longitudinal contact of the assembling monomers is in a
head-to-tail fashion. The other crucial eukaryotic cytoskeletal
protein - actin - also shows a distinct asymmetry. It forms
double-helical thin filaments composed of two strands. Within these,
actin assembles in a head-to-tail manner, similar to its bacterial
homologues [Michie and Lowe (2006)]. Indeed, another association
between the cytoskeletal network and percolation systems (Traverso
2005), recalls the long-range connectivity of magnetism (e.g.,
magnetic percolation clusters forming fractal networks (Itoh et al.
2006)). Again, the diamagnetic anisotropy of planar peptide bonds
permits their oriented self-assembly in a magnetic field, seen for
fibrous biostructures (Torbet and Ronziere 1984 plus ref).

% 3.4
\subsection{Reversible gel-sol transitions}
A far cry from organelles floating in sacs, the cytoplasm appears to
have rich structure irrespective of species, with increasingly
reported associations of mobile proteins with defined, albeit
transient, locations (Harold 2005). Again, 'site-dipoles' have been
proposed for resolving the apparent contradiction between the
seemingly random molecular movements and the correlated orientations
in assemblies. Thus the co-operativity among water molecules
occupying the site-dipole field surrounding a solute in MD
simulations, manifested in coherent patterns ($\sim$14A$^o$) that
lasted about 300ps, even as individual molecules randomly moving in
and out of the sites, rapidly lost their orientational memory [Higo
et al (2001)]. Indeed, the cell is viewed as a gel; reversible
gel-sol phase transitions underlie its dichotomy that can be
accessed via subtle environmental variations leading to finite
structural changes [Trevors and Pollack (2005)]. Like hydrated
cross-linked polymer gels, the cytoplasm thus exhibits excluded
volume effects and sizeable electrical potentials. Biomolecules like
proteins and ions play a critical role in structuring of
intracellular water (Chaplin 2004; 2006). This capacity to lie on
the border between liquid and gel states underlies life's ability to
make the most of fluidity of the liquid state as well as long range
order of the more solid gel phase, enabling self-assembly of
soft-matter. Now, in the origins of life, unlike a chemically bonded
thermally formed gel, a magnetic gel has the potential of reverting
back to its colloidal components just like $colloid-gel$ transitions
pointed out in living systems (Trevors and Pollack 2005).

% 3.5
\subsection{Reversible interactions; weak bonds}
The sensitivity of biomolecular machines to thermal noise is a
rather intriguing phenomenon. And, they have evidently learnt to
harness these, thanks to the continuous nature of the energy
landscape connecting different states. Again the interconvertibility
between different states is permitted due to the use of weak
interactions (Van der Waal's, H-bonds, hydrophobic, etc), used for
their temporary maintainance. Significantly, the Berry's phase-like
periodic cycles (Astumian 2007) shown by bio-molecular motors reveal
different trajectories for two half cycles (with different binding
capacities in forward and backward directions), that can be
understood in terms of their internal degrees of freedom. How could
such complexity of biological macromolecules have arisen from simple
matter, e.g. small molecules with a few discreet energy states,
present at the dawn of Life? This is since these very features
underlie the efficiency of biological machines that are being
increasingly viewed as microscopic systems governed by the
fluctuation-dissipation theorem (in the linear regime). The
variations in total Gaussian-distributed energy of a macroscopic
system with N particles, relative to the average value, are of the
order $N^{-1/2}$. Thus fluctuations would be negligible for
macroscopic systems, but they would be relevant for microscopic
ones, and also when the total energy of the system is $\sim k_B T$.

Next, in small systems in equilibrium or non-equilibrium steady
states, the behaviour remains unchanged in time, although a constant
input of energy is required for the latter, operating away from
equilibrium. No net heat transfer occurs in the former, with equal
probabilities of absorbing/releasing heat from bath. However, the
probability ratio differs from one for nonequilibrium steady state
systems that dissipate heat on the average. And heat, being an
extensive quantity, the probability of its absorption becomes
exponentially smaller with increasing system size. On the other
hand, for microscopic systems like bio-molecular machines driven by
rectified thermal fluctuations, this Maxwell-Demon-like probability
can be significant (Bustamante et al 2005). This has been very
succinctly phrased in a recent review (Haw 2007) as follows: "These
engines have one foot in the equilibrium camp and another in the
world of fluctuations and non-equilibrium". Indeed, Jarynski (1997)
showed that the average of the exponential of the energy of a
microscopic system, pulled quickly away from equilibrium (instead of
the simple average) works out to have the same value as the
equilibrium energy change corresponding to a slow version of the
same. This prediction was experimentally verified by Bustamante et
al (2005), where the result remained unaffected upon changing the
applied shearing force. In this proposal, diffusion of tiny
particles driven by non-equilibrium energy, via infinitesimal
changes in their relative orientations through the magnetically
ordered phase in a close-to-equilibrium manner shows the possibility
of controlled dynamics analogous to ATP-driven bio-molecular motors
(see Mitra-Delmotte and Mitra (2009)). Here the source of
non-equilibrium energy is none other than the gentle gradient of
flux lines thanks to a rock magnetic field (non-homogeneous).

% 3.6
\subsection{Kinetic barriers; records of constraints via hysteresis}
A major difference in the dynamics of life's processes lies in the
shift of the role of thermodynamics from a directing force in
regular chemical reactions to one of supporting the kinetics (Pross
2005). In fact, biology teems with examples of chemical reactions
that are thermodynamically allowed but await help for going across
the kinetic barrier-an intermediate state requiring energy of
activation ($E_a$), with the reaction rate primarily dictated by the
Boltzmann factor (exp(-$E_a$/kT)). Catalytic enzymes bring down the
barrier by enabling the appropriate relative positioning of the
reactants for reaction to occur. In the Hadean, rigid mineral
crystals could have acted likewise although it is difficult to see
how entire metabolic cycles of disparate reactions could have been
catalyzed on the same surface (Orgel 2000). On the other hand,
field-energy transfer through a network of magnetic templates within
the structured phase (Mitra-Delmotte and Mitra 2009) offers an
alternative scenario for enabling the juxtaposition required for not
only one but an array of reactions, by harnessing thermal
fluctuations to orient substrates diffusing into and binding to the
templates (c.f. Patel's (2006) oscillator inspired catalytic
mechanism for each reaction, see also Sect.3.9).

Note that Pattee's perception of life-dynamics arising out of an
irreducible 'whole'-internal interpretation of time-independent
symbolic codes (DNA) by their dynamical functional self-expressed
constraints (proteins)-- neatly subsumes the debate of which branch
of life-the metabolic or the replicator-first made its appearance in
the origins. Briefly, it may be recalled that constraints create
specific conditions for execution of physical laws in the dynamical
system they cause their local action thanks to frozen degrees of
freedom in their material structures. Their formation, in turn
depends on records or memory-like preserved constraining
configurations, e.g. the dislocation of a growing crystal. And,
although these do not form as a consequence of the dynamics of the
system in which they function (giving them an elevated `status'),
they can govern some dynamical events, by switching on-off in a
specific manner. In one scenario of such 'entangled' emergence of
symbols and metabolism -- a 'protometabolic' system-- where the
information specifying the network is distributed in its
organization (a membrane-enclosed recursive network of component
production) evolves to a self-interpreted genome via a stage
dependent on $non-symbolic$ records. This is crucially dependent on
the latter's ability to act at two levels: as a memory to be
expressed and as a way to express this memory (Etxeberria and Moreno
2001). Now, the phenomenon of hysteresis in magnetic materials,
provides a natural mechanism for the emergence of constraints in a
magnetically ordered system. For example, for reactions catalyzed on
the magnetic templates (Mitra-Delmotte and Mitra 2009) as above, the
imprint of the bound product in terms of altered orientations of the
template particles, would itself provide an 'observing' mechanism
for `recording' (the product of) the reaction.

% 3.7
\subsection{Self-reproduction; pre- bio-molecular motors}
Not only genetic information but entire progeny are modelled on the
'parent template' that provides the precise spatial information for
element organization and patterns, at different levels of the
intricately connected hierarchy [Harold (2005)]. Living systems use
diverse modes for copying patterns of information : nucleic acids
follow a template basis for assembly, membranes grow by extension of
existing ones, entire structures can duplicate (e.g. spindle pole
body or the dividing cell as a whole), etc. (Harold 2005). Now, in
contrast to the growth of mineral crystals (in traditional
origin-of-life models) restricted to a growing surface,
field-assisted alignment of diffusing tiny particles would occur
through the 'layers' of the accreted assembly, leading to its
inflation.

\begin{figure}
\centering
\includegraphics[width=120mm,height=100mm]{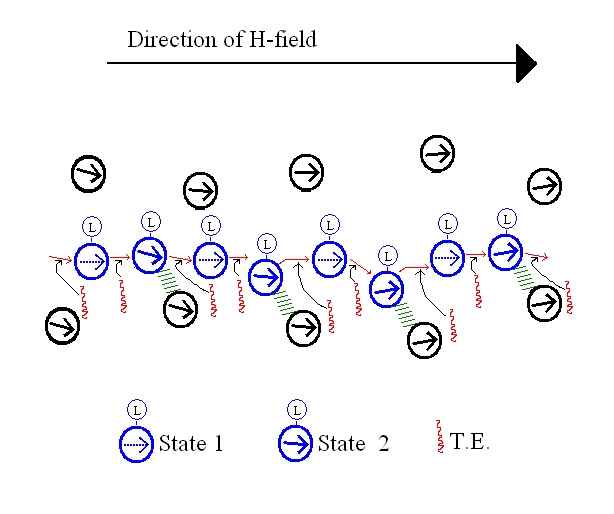}
\caption{Directed interactive diffusion of S-PP through MSP (with
parallel correlations). MSP represented in black;State 1/ State 2:
lower/higher template-affinity states of the ligand (L) -bound S-PP,
in blue; green lines signify alignment in State 2; T.E. or thermal
energy from bath; rock H-field direction indicated on top of figure,
see text}
\end{figure}

\par
Next, for ratchet-like effects, consider super-paramagnetic
ligand-bound particles (Milner-White and Russell 2005, 2008),
diffusing through the structured phase in an oriented manner as a
consequence of gentle change in flux lines, assuming that magnetic
rocks would have provided a non-homogeneous field. Now, two changes
are expected upon ligand binding: lowering of both rotational
freedom and coercivity (Vestal 2004) on the ligand-bound end. Thus,
while unconstrained rotation of ligand-free particles enables
alignment and propagation of the `information' in the magnetic
dipole-ordered assembly ('reproduction' as above), ligand-binding
aids diffusive passage. The constituents of the structured phase-
magnetically networked dipoles - are expected to locally perturb the
H-field 'seen' by the aligning and diffusing particles, moving
through its layers-the 'templates' (Figure 1). Thus alignment to
consequent template-partners would be alternated by dissociation
from the template, in cycles. Infinitesimal steps leading to these
altered states would require $\sim k_B T$, hence could be
facilitated by Brownian hits. This way the main features of today's
biological molecular motors: a non-equilibrium force applied
close-to-equilibrium that could reign-in Brownian noise, plus
asymmetry (via an H-field gradient), can be recovered
(Mitra-Delmotte and Mitra 2009). For, no macroscopic thermal
gradient runs these engines. Recall that a `thermal gradient' was
proposed by Feynman to circumvent the idea of 'biased' Brownian
motion (based on structural anisotropy alone) which, despite a right
magnitude for driving nano-sized particles (Phillips and Quake 2006)
is otherwise forbidden by the Second Law of Thermodynamics. The
evolution of these motors can perhaps be understood in terms of
non-magnetic `replacements' allowing the exit of such a magnetic
system from its geological confines (Mitra-Delmotte and Mitra 2009).
Indeed, diffusing super-paramagnetic units through a viscous medium
(due to inter-particle magnetic dipolar forces) have a striking
parallel to the directed movement of bio-molecular motors (in the
translational, transcriptional, cytoskeletal assemblies) on
aperiodic intracellular surfaces that indicate an invariant
topological theme for a ratchet mechanism, namely, movement of a
cargo loaded element on a template (representing a varying
potential) that harvests thermal fluctuations for dissociating its
bound state and spends energy for conformationally controlled
directed binding, or an ionic gradient for direction (Astumian
1997).

% 3.8
\subsection{Pre-RNA world; transfer reactions; optical activity}
Both magnetic templates as well as the particles (free or chemical
ligated) diffusing through the phase are part of a magnetically
connected network, and therefore seem to have the potential to
naturally provide topological correspondences to a variety of
biophenomena. For example, in the proposed RNA world, RNA played the
roles of \textit{both} DNA and protein - let's call them
RNA-sequential and RNA-structural, respectively. Evidently, nature
designed DNA for packaging information efficiently, satisfying
Shannon's maximum entropy requirement (no correlations across
sequences). This leads to the 'chicken-egg' conundrum, as the
largely random sequential information encoded in DNA is correlated
via RNA with the high degree of stereo-chemical information in
proteins. Now in contrast to hard periodic crystal lattices forged
with chemical bonds, confining physical forces in an accreted
ensemble gives a natural access to aperiodic surfaces (Breivik 2001,
see Sect.2.2). We therefore point out that RNA-sequential has
obvious parallels with aperiodic layers of a magnetically structured
phase hosting directed diffusion of ligand-bound super-paramagnetic
particles (above). These very 'templates' seem like a primitive
translational machinery, where W\"achtersh\"auser's (1988) `bucket
brigade-like' transfer reactions carried out by oriented particles
play the key adaptor roles \textit{a la} transfer RNAs - the
directed diffusion of the particle on an aperiodically packed
surface with no correlations (RNA-sequential-like), with the other,
ligand-bound to compounds rich in structural information. This
`magnetic letters-like' scenario bears a striking resemblance to the
tRNA's bringing the amino acids together for stringing them up on
the basis of the sequential information inscribed in the mRNA
template. And the maintenance of similar orientation, during
diffusive migration (depending upon the gradient of flux lines
cutting through the magnetically structured phase, i.e.
forward/backward from N to S or S to N; see above) offers a natural
mechanism for generating optical activity through symmetry-breaking.
This is because the solid-phase-like $arrangement$ of ligands, from
a racemic mixture (and bound to diffusing-super-paramagnetic
particles oriented to the magnetic-rock field), would take place in
the $limited$ space between densely packed magnetic layers/templates
(c.f. Viedma et al 2008; McBride and Tully 2008). And, in the
transfer reactions, this directional asymmetry of transport of an
oriented dipole due to a non-homogeneous external field has the
potential to push the balance in favour of bond formation between
juxtaposed activated units having the same chirality close to the
ligand-binding site. This is further aided by the space constraints
of such intra-layer activity, where the optical activity of the
first-bound unit (the symmetry-breaking choice) would set the
preferences for those of the subsequently selected ones.

% 3.9
\subsection{The potential for a quantum-leap to life}
These non-trivial correspondences between biological and magnetic
phenomena in general and topological correspondences to our proposal
in particular, prompt us to push this interface between these
apparently unrelated disciplines, to wonder why the
functional-approach-based selection of chemical molecules (where
changes are largely due to environmental fluctuations) would not
have started from a magnetic scaffold defining and dictating these
functional/contextual requirements? Indeed, the orientation of each
(particle) moment can be viewed as an interpreting gauge of its
composite environment-external field (rocks); neighbouring particle
moments; thermal fluctuations. It offers a 'route' for capturing a
"stable internal symbolic representation of the environment" to
borrow a phrase from Hoffmeyer (2001). So could there have been a
possible role of magnetism in endowing a system with constraints,
non-creativity, no goals, with the potential to jump to a state with
formal processes of controls, learning and instructions, creativity
(as in the extended version of Pattee's work drawn by Abel (2008) -
life as a bona-fide natural programmer), thus empowering the initial
conditions for this leap? In this connection it may be recalled
that, using the metaphor of an arch of stones, Cairns-Smith had
proposed that the scaffold paving the way for 'organic takeover'
(the `arch') may well have been provided by clay minerals that were
eventually disposed off. Indeed, this idea finds a sort of echo in
the suggestion of Patel (2002), viz., the choice of carbon with its
tetrahedral geometry provide the simplest discretization of the
fundamental operations of translation and rotation needed for
processing structural information. (Rotations in 3-D are $not$
commutative, a fact of crucial importance in representing structural
information; in mathematical jargon this goes by the name of the
SU(2) group of Pauli matrices/quaternions). Of course 'replacements'
via quantum searches could have well have been biopolymers with
capacity for classical searches that would have been more robust
against decoherence (c.f. the classical wave algorithm proposed by
Patel (2006)). According to Patel, vibrations and rotations of
molecules being harmonic oscillator modes, the catalyst like a mega
oscillator can focus the energy of many modes onto the reactant
awaiting activation.

This brings us to an important feature accessible via magnetism,
viz., a sound entry point for quantum processing. The Matsuno group
(2001) has reported the coherent alignment of induced magnetic
dipoles in ATP-activated actomyosin complexes that was maintained
over the entire filament even in the presence of thermal agitations
causing rapid decoherence. The energy of the dipole-dipole
interaction per monomeric unit of $1.1\times 10^{-22}$ Joule was
found to be far below the thermal energy per degree of freedom at
room temperature. This also can be extended to magnetically aligned
particles in a natural way. Work is currently in progress regarding
the role of a magnetic environment in aiding coherence. This matches
with Abel's (2009) observation, "…an inanimate environment has no
ability to program for a $potential$ function that does not yet
exist. Yet selection for potential function is exactly what genetic
programming requires". Thus Abel projects life as a bona fide
programming system with discretized instructions. Now, the
infinitesimal orientational changes of particles (associated
moments) diffusing through the layers of the assembly
(Mitra-Delmotte and Mitra 2009) offer yet another occasion for
discretization of operations required for processing structural
information, e.g. choice of carbon polymers (see above). Indeed, the
implications of a ferrofluid network as an analogue device can be
seen in the recent simulations by the Korenivski group (Ban and
Korenivski 2006; Palm and Korenivski 2009). We therefore suggest
that these magnetic nano-particle assemblies, could have been the
$soft-magnetic-matter$ version of Cairns-Smith's mineral scaffold
that was replaced by organic matter.

Again, one can find an example of discretization in the biological
currency ATP, providing energy for coupling to biochemical
reactions. Furthermore, in what is seen as a temperature lowering
mechanism enabling molecular motors to act as heat engines, Matsuno
and Paton (2000) describe the gradual release of energy stored in
ATP by actomyosin ATPase, in a sequence of quanta $E_m $ over time
intervals of $\Delta t_{m}$. This underlies the huge order of
magnitude discrepancy between the observed time interval of
hydrolysis of 1 molecule of $ATP \sim 10^{-2}$ sec,  and that
calculated by considering energy release of $E = 5\times 10^{-3}$
erg (7kcal/mol) from a singly emitted quantum, or $\hbar / E \approx
2 \times 10^{-15}$. The obtained values of $E_m \sim 2.2 \times
10^{-19}$ erg and $\Delta t_{m} \approx  4.5 \times 10^{-9}$ sec
indicates therefore $2.2 \times 10 ^{6}$ number of coherent energy
quanta release during one cycle of energy release from a single ATP
molecule. In Kelvin scale, each energy quantum $E_m$ amounts to $1.6
\times 10^ {-3}$ K associated with the actomyosin complex. Here too
we find that a mechanism enabling interchange between the a system's
environmental temperature and its own entropy is provided by the
(anistropic) magnetocaloric effect (MCE) (Tishin and Spichkin 2003),
which is the property of some magnetic materials to heat up when
placed in an H-field and cool down when they are removed
(adiabatic). In fact, the heat capacity at the nano-scale turns out
to be a few-fold higher than that of bulk systems, thanks to MCE
(Korolev et al 2008). We have suggested (Mitra-Delmotte and Mitra
2009) that the exit of the `magnetic ancestor' from the confines of
its magnetic environment may have been enabled upon coupling of its
envisaged dynamics associated with changes in gradient of flux
lines, instead with ATP-the universal biological currency (see Sect.
5.4; also Sects.2.4, 3.5, 3.7).

In this scenario, biological phenomena with similarity to magnetic
ones could be considered as `distant cousins' of their `non-living'
counterparts. Thus even quantum processing is viewed as a legacy and
not a product of adaptive evolution (Doll and Finke 2003). Note that
magnetic ordering may stem from unpaired p - electron systems
(Ohldag et al. 2007) (not just 3d, 4f!). The 'substitutes', despite
increasing complexity, would need to pass-on the legacy of
multi-dimensional properties of the Ancestor possessed, especially
phase information, e.g. DNA has positional information, with
possible phase signatures in its helical structure (Kwon 2007).

% 4.0
\section{Framboids and the mineral greigite}
We shall now seek to expand the potential of mineral crystal
theories, by looking for minerals that can enable magnetic effects,
such as those outlined above. This brings us to framboids (Wilkins
and Barnes 1997, see below) as these dynamically ordered
terrestrial/extraterrestrial, microcrystal composites formed by
structurally $different$ materials show the control of packing by
$physical$ forces.

% 4.1
\subsection{Framboids; importance of physical properties}
In framboids, named after their framboise/raspberry-like patterns,
nucleation of clusters is followed by growth of individual nuclei
into microcrystals. They have been defined as microscopic spheroidal
to sub-spheroidal clusters of equidimensional and equimorphic
microcrystals which suggest a homogenous nucleation of the initial
microcrystals. Other than the spherical framboids, a highly ordered
icosahedral type has been reported where this packing is maintained
in its internal structure. The formational environment is evidently
critical for the packing in these varied forms. As pointed out by
Ohfuji and Akai (2002), D/d ratios of framboids (framboid diameter D
and microcrystal diameter d) dominated by irregular or loosely
packed cubic-cuboidal microcrystals are low compared to high
corresponding values observed for those composed of ordered densely
packed octahedral microcrystals. The narrow distribution of sizes
and uniform growth of thousands of crystals in framboids within a
short time interval was attributed to a regulated balance between
rates of nucleation and of crystal growth, as in the La Mer and
Dinegar model (1950). Furthermore, the nucleation of a
supersaturated solution by the first-formed crystal triggers the
separation of many crystals of the same size. This liquid-solid-like
phase transition is dependent on packing considerations of
hard-sphere-like microcrystals, whose ordering is an outcome of the
interplay of close-packing and repulsive forces (see Sawlowicz
2000).
\par
As noted by Sawlowicz (2000), the framboidal texture is seen in a
number of different minerals other than pyrite, i.e. copper and zinc
sulphides, greigite, magnetite, magnesioferrite, hematite, goethite,
garnet, dolomite, opal, and even in phosphoric derivatives of
allophane. This suggests a similar mechanism of formation, despite
the structural differences. Studying their presence in sedimentary
environments, Sawlowicz (1993) found pyrite framboids to be
hierarchially structured over three size-scales: microframboids, to
framboids, to polyframboids. And since spheroidal microframboids are
formed of equant nanocrystals, he suggested (1993, 2000) the
formation of nano-framboids, comprising microcluster aggregations
($\sim 100 atoms$), by analogy with the 3-scale framboidal
hierarchy. His observations leading to a proposed formation
mechanism center around the key role of the colloid-gel phase
leading to the fractal forms. Interestingly, exclusion of organic
compounds, were found to lead to simple framboids via an aggregation
mechanism while experiments with organic substance stabilized
gel-droplets, framboids formed by $particulation$. This latter route
is seen as important for generating the $fractal$ complexity.
Similar scale free framboids of greigite that is ferrimagnetic
(next), have also been documented (Preisinger and Aslanian 2004).

% 4.2
\subsection{Framboidal greigite}
In framboids reported in sedimentary rocks more than 11,000 years
old [Roberts and Turner (1993)], the central parts of the weakly
magnetized framboids were found to have greigite microcrystals.
Sections from these, show that the pentagonal arrangement comprise a
central pentagonal domain with its sides connected to five
rectangular/trapezoid-like regions which are in turn connected via
fanshaped domains. The arrangement pattern of these densely packed
octahedral microcrystals linked edge to edge is 'lattice-like'
(space filled) in the rectangular domains, whereas in the triangular
domains the triangles are formed by the (111) faces of the
octahedral microcrystals and the voids between them. Thus within
these domains the individual faces of the microcrystals do not make
any contact. The icosahedral form is seen as generated by stacking
twenty tetrahedral sectors packed on three faces out of four, and
connected by their apexes at the centre. Generally acknowledged as
dynamically stable, this form is known to have six 5-fold axes at
each apex, and ten 3-fold axes at each face, as can be seen in a
number of naturally occuring structures from microclusters like
fullerene to some viruses (Ohfuji and Akai 2002). Further, in an
investigation of apparent biologically induced mineralization by
symbiotically associating bacterial and archaeal species, framboidal
greigites have been obtained from Black Sea sediments that are
ordered clusters of octahedral crystals comprising $Fe_3S_4$-spinels
(Essentially cubic where sulphur forms a fcc lattice with 32 atoms
in the unit cell, and Fe occupies 1/8 of the tetrahedral and 1/2 of
the octahedral sites). Their size is restrained by their icosahedral
symmetry and under greater pressures at depths of 200m, the
diameters are mostly $\sim$ (2.1, 4.2, 6.3 or 8.4) $\mu$, with the
two intermediate ones predominating. The smallest of these are
formed from 20 octahedral crystals (0.35 $\mu$) positioned at the
apexes of an icosahedron and surrounding a 0.5 $\mu$ diameter
vacancy that give rise to 12 pentagonal depressions on the outside.
Nested structures building up from this smallest one lead to the
higher sized clusters (Preisinger and Aslanian 2004). Sub-spheroidal
pyrite-framboids, due to curved polyhedron-like outer facets,
probably reflect an internal icosahedral microcrystal organisation
(Ohfuji and Akai 2002), which are classically forbidden
crystallographic symmetries (Ohfuji and Rickard 2005).

% 4.3
\subsection{Magnetic interactions}
Magnetic interactions turned out to have an overwhelming influence
when Wilkin and Barnes (1997) included them in the standard DLVO
treatment for interacting colloidal particles that considers
attractive van der Waals and double-layer repulsive interactions,
for modeling framboidal pyrite formation. This is based on the
alignment of precursor greigite, under the influence of the weak
geo-magnetic field that would help overcome the thermal energy of
particles above a critical size. Ferrimagnetic greigite has a
saturation magnetization value $M_{sat}$ at 298K ranging between 110
and 130 $kA/m$. On the basis of microscopic observations by Hoffmann
(1992) of natural greigite crystals, $<$ $\mu$ meter - sized
greigite can be roughly taken as single-domain particles. Assuming a
spherical geometry, the critical grain diameter of constituent
crystallites comprising the framboid interior $d_c = 2a$, where $a >
1$, is given by
\begin{equation}\label{1.1}
d_c = (6 k_BT /\mu_0 \pi M_{sat}|H|)^{1/3}
\end{equation}
This result can be obtained from the inequality $ W_{WB} > k_B T$
where we $define$ $ W_{WB} \equiv \mu_0 M_{sat} V H $. Here $k_B$ is
Boltzmann's constant and $\mu_0$ the permeability of vacuum. When
aligned parallel to weak geomagnetic field ($\sim$ 70$\mu$T), $d_c$
= 0.1 $\mu$m. Although framboids can form in varied environments and
by other mechanisms (see Sawlowicz 2000; Ohfuji and Rickard 2005),
this magnetic greigite-precursor mechanism can operate only upto
temperatures of 200 $^{\circ}$ C (Wilkin and Barnes, 1997), eg.
sediments, in natural waters. Also, as pointed out by Wilkin and
Barnes (1997), the effect of weak fields leads to spherical
structures in ferrofluids (Sect. 2.3) in contrast to aspect ratios
approaching infinity in strong fields. They also noted the role of
turbulence in facilitating the interplay of opposing interactions.

% 4.4
\subsection{Dynamic ordering; phyllotaxis; quasiperiodicity}
A characteristic pattern of icosahedral framboids - octahedral
microcrystals, large D/d ratio - has been attributed to a high
initial nucleation rate and low growth rate of microcrystals (Ohfuji
and Akai 2002; Ohfuji and Rickard 2005). According to Sawlowicz
(2000) the interplay of surface-minimizing forces with repulsive
interactions lead to close-packed framboids, tending to polyhedrons.
And, this is a ramification of anastrophic supramolecular
organization, with its far-from-equilibrium conditions. Sure enough,
the framboid morphology is strongly remniscent of the ubiquitous
phenomena of Phyllotaxis, from subnano to cosmological scales
(Levitov 1991; Adler 1997; Dunlap 1997): Repulsive magnetic dipoles,
galactic structures, biostructures, from the molecular (proteins,
DNA) to macroscopic levels (myriad  marine forms), proportions in
morphological and branching patterns (Dunlap 1997), Benard
convection cells, stress-driven self-assembly, bunched crystalline
ion beams, atmospheric flows, and flux lattices in layered
superconductors. Phyllotactic patterns are produced when the
sequential accretion/deposition or appearance/growth of elements, is
governed by an energy-minimized optimization of the main opposing
forces: largest available space vs repulsive interactions. And in
magnetically accreted greigite framboids (Wilkin and Barnes 1997)
too, a similar interplay of conflicting forces, leads to
raspberry-like phyllotactic patterns. This dynamic ordering via
accretion of magnetic crystals in the face of short-range repulsive
forces does contrast with the build-up of a conventional infinite
crystalline lattice, where the nuclear surface acts as a template
for copying a unit cell via local interactions. Rather, it is
analogous to a scenario at nano-scales--one associated with the
aperiodic, long range order of systems known to form $quasicrystals$
whose growth occurs by accretion of pre-formed clusters in the
liquid state by the growing nucleus (Keys and Glotzer 2007). Now,
the relevance to greigite concerns its natural preference for such
order as evidenced from observations of nested scale-free
icosahedral greigite framboids (Preisinger and Aslanian 2004). These
observations are intriguing in view of the known links between
phyllotactic patterns and quasiperiodic phases. For instance the
predominance of edge-to-edge contacts between microcrystals
comprising icosahedral greigite framboids (Roberts and Turner 1993;
Ohfuji and Akai 2002) limits possible $conduction$ pathways.

% 4.5
\subsection{Magnetic assemblies in the laboratory; long-range order?}
Some insights into the above natural assemblies are offered by
synthetic ones driven via a different route of evaporation (Ahniyaz
et al 2007; Sun et al 2000; Cheon et al 2006), also one under
hydrothermal conditions (Wu et al 2005). Apart from external-field
control, other physical properties of nano-constituents:
crystalline/colloidal state, geometry, susceptibility, coatings,
etc, are important criteria for clustering patterns (Pileni 2003;
Lalatonne et al 2004). Next, in soft condensed matter studies,
varied and unusual polyhedra have been seen in packing sequences of
colloidal polystyrene microspheres, illustrating how certain
symmetries, including five-fold rotational symmetry, can arise
solely from compression and packing constraints. These can explained
by the use of a minimization principle - that of the second moment
of mass distribution wrt the center of mass ($\Sigma_i m_i x_i^2 $),
instead of the conventional volume ( $\sim r^3$) --optimizes the
packing (Manoharan et al 2003; review in Yethiraj 2007). Again, the
$route$ to formation is another important aspect of assembly; there
is no possibility of an internal sphere upon collapse in this
evaporation-driven system that starts from spherically packed
particles bound to a continuous and smooth (2D) surface, i.e. the
droplet interface. This route would not apply to particles
compressed via magnetic dipolar forces as in scale-free greigite
framboids, which is more like a problem of packing spheres not only
on the surface of a sphere (2d-space), but rather $into$ a finite 3D
space, as in some compounds, alloys, quasicrystals that have long
range order without periodicity. Recall that framboidal texture
comes via optimized packing of microcrystals (see large $D/d$
ratios, Sect. 4.1). That structurally different materials form
framboids (Sect. 4.1) also reveal the important role of the
colloidal state where physical properties can be accessed, in
contrast to the strong influence of chemical properties for packing
in (periodic) crystals. An understanding of icosahedral geometry in
scale-free greigite framboids can be had from a study of tesselation
of spheres (number N; radius a) packed on the surface of a large
sphere (radius R). This shows that energy minimization would lead to
buckling into icosahedral forms, considering only small R/a ratios,
as $N \sim (R/a){^2}$ (Nelson 2003). This in turn could bring in
geometrical frustrations but studies on icosahedral magnetic
quasicrystals (Lifshitz 1998) show that geometrical constraints do
$not$ rule out the possibility of long range magnetic order.

\par
Thus we find that in the mineral inorganic world too, superimposed
physical interactions can dictate assembly organization.
Furthermore, it is significant that greigite, which is known to
undergo accretion due to magnetic forces (Wilkins and Barnes 1997)
and also has a natural propensity for framboid formation (Ohfuji and
Akai 2002; nested forms in Preisinger and Aslanian 2004), is also
strongly suspected for its `metabolic' potential (next).

% 5.0
\section{Mound scenario of Russell et al and greigite}
In fact, the search for greigite forming on the Hadean Ocean floor
led us to the colloidal environment setting of Russell and coworkers
where greigite forms across gradients and that leads to a
metabolically enriched scenario (next).

% 5.1
\subsection{Mound scenario of Russell et al}
The colloidal environment-based proposal of Russell et al (1994)
envisages Life as having emerged in moderate temperature
hydrothermal systems, such as mild alkaline seepage springs. Water
percolating down through cracks in the hot ocean crusts reacted
exothermically with ferrous iron minerals, and returned in
convective updrafts infused with H$_{2}$,  NH$_{3}$, HCOO$^{-}$,
HS$^{-}$, CH$_3^{-}$; this fluid (pH $\sim$ 10 $\leq$ 120$^{\circ}$
C) exhaled  into CO$_{2}$, Fe$^{2+}$ bearing ocean waters (pH $\sim$
5.5 $\leq$ 20$^{\circ}$ C) (Russell and Arndt 2005). The interface
evolved gradually from a colloidal FeS barrier to a single membrane
and thence to more precipitating barriers of FeS gel membranes.
Since fluids in alkaline hydrothermal environments contain very
little hydrogen sulphide, the entry of bisulphide, likely to have
been carried in alkaline solution on occasions where the solution
met sulphides at depth (Russell and Hall 2009), was controlled. This
was perhaps important for a gradual build-up of scale-free clusters
leading to the envisaged gel-environment. (As pointed out by
Sawlowicz (2000) colloids often form more readily in dilute
solutions -- suspension as a sol-- than in concentrated ones where
heavy precipitates are likely to form). These barriers controlled
the meeting of the two fluids, as they enclosed bubbles entrapping
the alkaline exhalate : an aggregate growing by hydrodynamic
inflation. The forced entry of buoyant seeps may have led to
chimney-like protrusions. Further, theoretical studies by Russell
and Hall (2006) show the potential of the alkaline hydrothermal
solution (expected to flow for at least 30,000 years) for dissolving
sulfhydryl ions from sulfides in the ocean crust. The reaction of
these with ferrous iron in the acidulous Hadean ocean (derived from
very hot springs, Russell and Hall 2006) is seen as having drawn a
secondary ocean current with the Fe$^{2+}$ toward the alkaline
spring as a result of entrainment (Martin et al 2008). Hence at the
growing front of the mound, the production of daughter bubbles by
budding would have been sustained by a constant supply of newly
precipitated FeS. Like cells, these mini FeS compartments protected
and concentrated the spectrum of energy-rich molecules, borne out by
harnessing important gradients across the mound (a true far-from
equilibrium system, driven by energy released from geodynamic
sources): redox, pH and thermal gradients for electron transfers,
primitive metabolism, and directed diffusion, respectively (Russell
and Arndt 2005). See also Rickard and Luther (2007) for an analysis
of the reducing power of FeS for synthesizing organics in this
proposed scenario.
\par

Experimental simulations of mound conditions using calculated
concentrations of ferrous iron and sulphide (20mmoles of each)
resulted in the formation of a simple membrane. Using solutions with
five to twenty-fold greater concentrations (to make up for their
build-up in geological time) generated compartmentalized structures,
shown in Fig.2 where the chambers and walls are $\sim$ 20 and 5
$\mu$, respectively. These have remarkable similarities to porous
ones in retrieved Irish orebodies, shown in Fig.3, which had
originally inspired the idea that the first compartments involved in
the emergence of life were of comparable structure (see Russell and
Hall 1997a; Russell 2007). In fact, even submarine mounds seen today
are invariably porous (Marteinsson et al. 2001; Kelley et al. 2005).
Also, the sulphide comprising what is now pyrite (FeS$_{2}$) in the
350 million year old submarine Irish deposits (Fig.3) was derived
through bacterial sulfate reduction  in somewhat alkaline and saline
seawater while the iron was contributed by exhaling acidic
hydothermal solutions. On mixing, mackinawite (Fe(Ni)S) and greigite
(Fe$_{5}$NiS$_{8}$) would have precipitated to form inorganic
membranes at the interface (Russell et al. 1994; Russell and Hall
1997a).

% 5.2
\subsection{Greigite formation from FeS}
Fig.2 shows laboratory simulated FeS compartments; the chambers and
walls are $\sim$ 20 and 5 $\mu$, respectively. According to Russell
et al. (2005), the permeable membranes likely comprise
(ferredoxin-like) greigite and mackinawite, and whose metal and
sulphide layers work for and against e$^-$ conduction, respectively.
An insight into this calls for a brief outline of iron sulphide
transformations under wet and moderate temperature conditions.
Amorphous mackinawite ($FeS_{(am)}$) is the first FeS phase formed
from aqueous S(-II) and Fe(II) at ambient temperatures, apparently
via two competing pathways governing the relative proportions of the
two end-member phase mixture. The long-range ordered phase with
bigger crystalline domain size and more compact lattice increases at
the cost of sheet-like precipitated aqueous FeS clusters (Wolthers
et al. 2003).

\begin{figure}[h]
\begin{minipage}[b]
{0.4\linewidth} \centering
\includegraphics[scale=0.75]{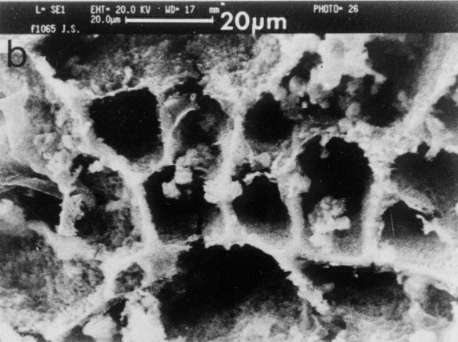}
\caption{FeS compartments} \label{fig:(a)}
\end{minipage}
\hspace{0.25cm}
\begin{minipage}[b]{0.6\linewidth}
\centering
\includegraphics[scale=0.7]{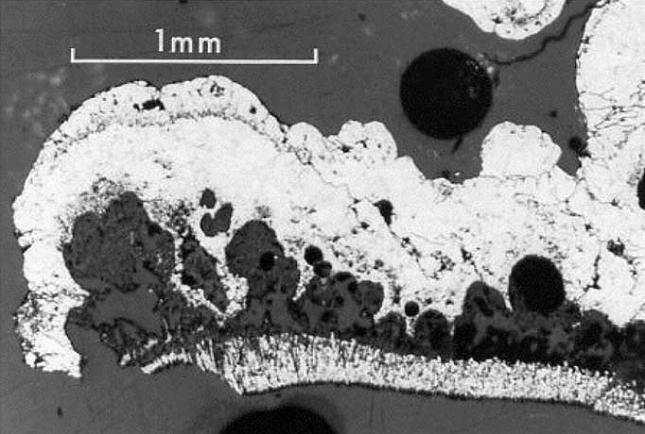}
\caption{FeS botryoids} \label{fig:(b)}
\end{minipage}
\scriptsize{FeS compartments and botryoids: Fig.\ref{fig:(a)}. SEM
photo of a freeze dried section showing FeS compartments formed on
injecting 0.5 M Na$_2$S solution into 0.5M of FeCl$_2$ (Russell and
Hall 1997). Reproduced with kind permission from M.J. Russell;
Fig.\ref{fig:(b)}. Polished cross-section of the Tynagh iron sulfide
botryoids. Kindly provided by M.J. Russell, see text for details.}
\end{figure}

Note that an FeS cluster can display two properties: 1) it can be
regarded as a  multinuclear complex (where instead of a central
atom, as in a complex, a system of bonds connects each atom directly
to its neighbours in the polyhedron); and 2) as an embryo since it
can develop to form the nucleus of the first condensed phase
(Rickard and Morse 2005). The formation of the latter gets initiated
by statistical fluctuations in the density of the initial parent
phase (e.g. due to supersaturation) and its growth is favoured by
the difference in chemical potentials between the parent and the new
phase. Reviewing aqueous FeS clusters in water environments, Rickard
and Morse (2005) suggested the enhanced stability of some
stoichiometries--stable magic number clusters-- from among the
apparent continuum of stoichiometries of aqueous FeS clusters. This
ranges from Fe$_{2}$S$_{2}$ to  Fe$_{150}$S$_{150}$, where the first
condensed phase (FeS$_{m}$, mackinawite) appears, with a size and
volume of 2 nm and 10 nm$^{3}$, respectively. Although molecular
Fe$_{2}$S$_{2}$ is similar in structure to crystalline mackinawite,
the Fe-Fe bond lengths and Fe-S-Fe bond angles are seen to approach
those of crystalline mackinawite, in tandem with increased size of
molecular FeS clusters. The decrease in degree of softness, or water
loss, can be gauged from the relative density increase over the
smallest Fe$_{2}$S$_{2}$ cluster ($\geq$ 10$^{6}$), as the structure
of hydrated clusters is believed to determine that of the first
condensed phase. X-ray diffraction of the first nano-precipitate
shows a (lattice expanded) tetragonal mackinawite structure. That
the data fit well with other independent estimates is ascribed to
the plate-like form of FeS$_{m}$. The quick transformation of
disordered mackinawite to the ordered form is followed by solid
state transformation to the more stable but structurally congruent
greigite, with a \textit{12 percent decrease in volume}, involving a
rearrangement of Fe atoms in a close-packed, cubic array of S atoms.
Further, trace amounts of aldehydes are believed to bind to the
$FeS_{(am)}$ surface, initiating Fe(II) oxidation (S( - II)
unaffected) ; they also prevent the  dissolution reaction,
$FeS_{(am)}$ to $FeS_{(aq)}$ (aqueous FeS complex), crucial for
pyrite formation (in absence of aldehyde, S(-II) oxidised, Fe(II)
unchanged), thus assisting in greigite formation at the cost of
pyrite (perhaps as in bacteria) (Rickard et al. 2001). Such a
solid-state transformation of amorphous mackinawite to greigite can
be extended to FeS clusters -- Rickard and Luther (2007) suggest the
possibility of organic ligands stabilizing aqueous Fe(III)-bearing
sulphide clusters, as seen in similar (greigite-like) cubane forms
in FeS proteins.  Importantly, FeS membranes formed in the
laboratory show a 20-40 fold increased durability on adding
abiogenic organics. Diffusion controlled reactions would slow down
with thickening of aging/hardening of membranes (Russell et al.
1994).

% 5.3
\subsection{The FeS Gel environment and framboids}
As noted by Russell et al. (1994), citing Kopelman (1989), gels lie
between liquid and solid states with self-similar clusters, fractal
on all scales (permitting diffusion control in heterogenous
reactions, ubiquitous in biosystems). They suggested (Russell et al
1989, 1990) the $nucleation$ of the FeS gel bubbles by iron sulphide
: in vitro simulations of iron sulphide chimneys demonstrated
formation of macroscopic spherical shells 1 to 20 mm across, while
on a microscopic scale spherical, ordered aggregates of
\textit{framboidal pyrite} about 5 micro meter in  diameter were
found in fossil hydrothermal chimneys (see Figure 3; Boyce et al.
1983; Boyce 1990; Larter et al. 1983) that seemed to have grown
inorganically from the spherical shells of FeS gel. These framboidal
sacks of periodic arrays within the extensive reactive surfaces per
unit volume of the chimneys, could have offered ideal experimental
culture chambers and flow reactors well poised for origin-of-life
experiments (Russell et al 1990). Indeed, framboids have long been
recognized for their fascinating features, prompting speculations on
their possible role in the origin of life, e.g. Sawlowicz (2000)
noted the bio-potential of constituent microcrystal surfaces,
presence of catalytic metals, fractal structures, to name some.

\begin{figure}[h]
\begin{minipage}[b]
{0.4\linewidth} \centering
\includegraphics[scale=0.5]{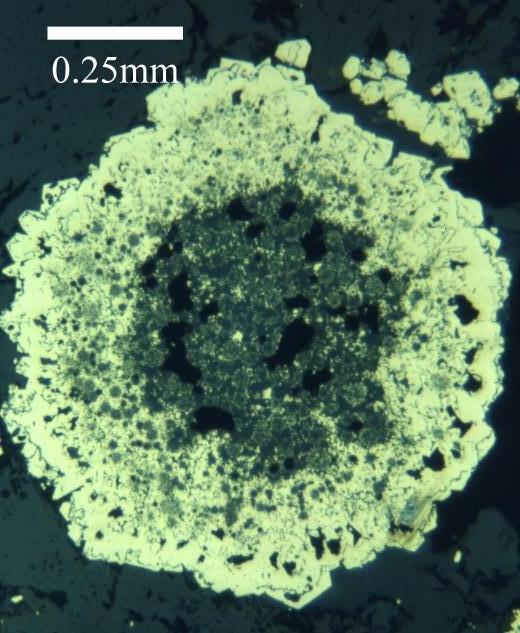}
\caption{Small Vent} \label{fig:(c)}
\end{minipage}
\hspace{0.3cm}
\begin{minipage}[b]{0.6\linewidth}
\centering
\includegraphics[scale=0.5]{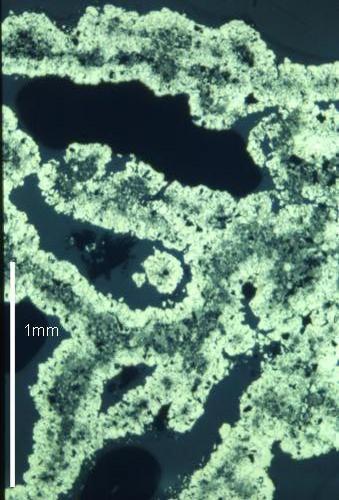}
\caption{Sheaves} \label{fig:(d)}
\end{minipage}
\scriptsize{Framboids in chimneys: Fig.\ref{fig:(c)}. Small pyrite
vent structure: Reflected ore microscopy of transverse section shows
a central area of empty black spaces plus (grey) fine framboidal
pyrite, and a fine euhedral authigenic rim surrounded by baryte,
with minor pyrite; Fig\ref{fig:(d)}. Sheaf system, formed from
coalescing rods of anastamosing microcrystalline pyrite. Black areas
are empty spaces; central regions are framboidal pyrite with an
exterior of crystalline pyrite. (Labelled pictures given by Dr.
Adrian Boyce are reproduced with his kind permission; Source: Boyce
et al. 1983; Boyce 1990: Exhalation, sedimentation and sulphur
isotope geochemistry of the Silvermines Zn + Pb + Ba deposits,
County Tipperary, Ireland; Boyce, Unpublished Ph.D. thesis,
University of Strathclyde, Glasgow).}
\end{figure}

\par

The above observations of magnetically accreted framboidal greigite
(Sect. 4.2) and possibility of framboid assembly in colloidal
environment lead us to think that super-paramagnetic greigite could
have formed magnetic assemblies (in the presence of magnetic rocks)
as starting self-reproducing systems, besides being a precursor for
nucleic acids, proteins, lipids, etc., that could have been chosen
as context-based replacements. This could be significant for it has
been long recognized that much of the path sketched from prebiotic
chemistry to the RNA world (a widely accepted hypothesis; see Orgel
2004) remains unchartered and for start points (see Shapiro 1999),
there are suggestions of 'physically' self-reproducing systems as
having preceded 'chemically-copying' self-replicators (Dyson 1999);
autocatalytic reactions (Kauffman 1993) and self-replicating
inorganic (Cairns-Smith 1982) or even a combination of organic and
inorganic (Orgel 1986) systems.

% 5.4
\subsection{Field estimate from W-B model; Motor-like dynamics}
We now come to the possibility of magnetic rocks which could further
expand the potential of the mound scenario, described above. The
associated H-field with rocks, needed for overcoming temperatures
$\sim$ 50C in the mound, is estimated by extrapolating the Wilkin
and Barnes (W-B) model (1997) for formation of framboidal pyrite via
the precursor greigite. When aligned parallel to weak geomagnetic
field ($\sim$ 70$\mu$T), it gives $d_c$ = 0.1 $\mu$m (see Sect.
4.3). Thus, a rock H-field for accreting 10nm sized particles would
have to be 1000-fold higher. This also is of the same order of
magnitude $\sim$ 10mT, seen for magnetite-based ferrofluids
(Odenbach 2004). For, the saturation magnetization of magnetite
($M_s$ = $4.46 \times 10^5 $ A/m) is about 3.5 times greater than
that of greigite; from this one expects proportionate values for the
fluid susceptibility of a corresponding greigite suspension,
building up slowly in the ocean waters (see above). Also, the
dipole-dipole interactions between negatively charged greigite
particles (as the pH is well above 3 under mound conditions (Wilkin
and Barnes 1997)) is likely to be aided by the screening effect due
to ionic strength of natural waters (Spitzer and Poolman 2009).

Now, as the geo-magnetic field did not even exist at $\sim$ 4.1-4.2
Ga (Hazen et al 2008) (whereas life is thought to have initiated at
$\sim$ 4.2-4.3 Ga (Russell and Hall 1997b; 2006), we look to local
sources for providing a magnetic field $\sim$ 50-100mT for enabling
accretion of newly forming greigite particles. (For example, the
present geomagnetic field strength is too weak to explain the
magnetization mechanism of lodestones). To that end, a plausible
candidate (c.f. Wasilewski and Kletetschka 1999) could be isothermal
remnant magnetism (acquired by lightning, impact, etc) in say,
meteoritic matter on its way to the Ocean floor. In fact, Ostro and
Russell have suggested plausible mechanisms for accumulation of
reducing meteoritic matter, around the base of the mound. Also,
unlike today's conditions, the primitive crust was still extremely
reducing when life is thought to have emerged (Righter et al 1997;
Russell et al 2003) making the presence of ferromagnetic matter a
likely event. Further reinforcement of the local H-field would occur
through the generation of magnetic minerals like magnetite and
awaruite (Dyment et al 1997, Schroeder et al 2002; Beard and
Hopkinson 2000) immediately beneath the mound due to
serpentinization of Ocean Floor peridotites   (for more details see
Mitra-Delmotte and Mitra 2009).

Here, magnetic rocks could have not only helped the accretion of
greigite particles, but also gentle changing flux due to
non-homogeneous field lines (expected from rocks) could have gently
moved incoming particles aligned to the field, i.e. in the same
orientation in either the forward or backward (N-S or S-N)
directions, depending upon their position in the structured phase,
and using thermal fluctuations to drive ratchet-like effects (see
Sect.3). At the same time, such a magnetic albeit locally confined
ancestor, maintained close-to-equilibrium, would also have the
potential for coupling with non-equilibrium energy sources (such as
pH or redox gradient) -the 'metabolic' wing of life-producing energy
rich molecules (Russell and Arndt 2005). This capacity of a
magnetically controlled system to couple to different gradients,
e.g. thermal (Baaske et al 2007), was also needed to pave the way
for complex energy transduction mechanisms. We have suggested
(Mitra-Delmotte and Mitra 2009) that the `innovative evolution' of a
bio- ratchet where coupling to non-equilibrium energy (in discrete
packets) from energy-rich molecules propelled close-to-equilibrium
dynamics (driven so far by a gentle H-field gradient), allowed the
exit of the Ancestor from its geological location for seeking out
gradient-rich niches elsewhere. This in turn would have led to a
progressively decreasing functional dependence on iron sulphide.
Nevertheless, the continued presence of magnetic elements (e.g.
structural roles) would offer a magnetic basis for the association
of its 'liberated' replacements as in the multicellular life
proposal (Davila et al 2007). The possibility of different `magnetic
soups' close to the mound also converges well with the suggestions
of Martin and Russell (2003), Koonin and Martin (2005), of an
initially confined universal ancestor diverging into replicating
systems, located separately on a single submarine seepage site (see
Sect. 5.1), en route to proto-branches of life. These
reproducer-turned replicators could navigate to different openings
where survival criteria would induce variations. The transfer of
regulatory powers to the genes is likely to have been slow but
progressive. In the pre-Mendelian era, there was more plasticity in
phenotype - genotype mapping, gradually taking on a one-to-one basis
with a decline in morphological plasticity - yet another
'robustness' enhancing strategy (Newman and Muller 2000).

% 5.5
\subsection{Enzyme clusters and natural violarite phases}
Note that the composition of iron sulphide clusters found in
enzymes, $Fe_5 Ni S_8$, lie between $Fe Ni_2 S_4$ and $Fe_3 S_4$.
Although a solid solution in this range has not been observed in
synthetic dry condition, high temperature experiments, it has been
observed in natural violarite (iron-nickel thiospinel) phases
(Vaughan and Craig 1985). More recently, the supergene oxidation of
pentlandite ($(Fe, Ni)_{9} S_{8}$) to violarite (includes extensions
from $Fe Ni_{2} S_{4}$ towards both $Fe_{3} S_{4}$ and $Ni_{3}
S_{4}$), was experimentally reproduced under mild hydrothermal
conditions (Tenailleau et al 2006). The results show the
$feasibility$ of high iron/nickel ratios in violarite forming under
reducing mound conditions, despite the suggested metastability of
these compositions from bonding models. Iron is believed to occur as
low spin $Fe^{2+}$ in $Fe Ni _{2} S_{4}$ that exhibits metallic,
Pauli paramagnetic behaviour. In contrast, the Mossbauer spectrum of
$Fe_{3} S_{4}$ is attributed to high-spin $Fe^{3+}$ in tetrahedral A
and Octahedral B sites and it's electronic structure from molecular
orbital calculations (Vaughan and Tossell 1981) reveal localized 3d
electrons with unpaired spins, coupled anti-ferromagnetically at
lower temperatures. According to Vaughan and Craig (1985), the
greater ionic character and larger number of electrons in
antibonding orbitals in $Fe_3 S_{4}$ relative to $Fe Ni_{2} S_{4}$,
could contribute to the instability of intermediate compositions,
despite their natural occurrence.

% 5.6
\subsection{Coherence: ferromagnetic-ferroelectric effects}
The quest for co-existing (in same or locally different subspaces)
ferroelectric effects reinforcing the coherent (`dispersive',
non-dissipative) effects of ferromagnetism arises out of interesting
present-day biological observations. Frohlich (1968, 1975) proposed
the emergence of a long range coherent state via alignment of
dipoles in cell membranes. Ordering of electric dipoles via
interactions between structured water and the interior of
microtubular cavities brings in a dynamic role of {\it
ferroelectricity} as a frequency-dependent dielectric-constant
$\epsilon(\omega)$, which gives a big dispersive (non-dissipative)
interaction (robust against thermal losses) for small values of
$\omega$ (since the factor $\epsilon(\omega)$ occurs in the
denominator of the corresponding interaction) (Mavromatos et al
1998). Apart from the importance of such coherent electric dipole
ordering alignment of actin monomers prior to ATP-activation, Hatori
et al (2001) report the coherent alignment of magnetic dipoles
induced along the filament, by the flow of protons released from ATP
molecules during their hydrolysis (basically a Maxwell displacement
current-like dynamical effect). But in contrast to the similar
nature of magnetic ordering mechanisms conferring ferromagnetism via
exchange interactions of predominantly localized magnetic moments, a
variety of ferroelectric ordering mechanisms exist for different
types of ferroelectrics, not all of which are well understood. In
fact, in materials their co-existence can range from being mutually
exclusive, such as due to incompatibility of d-electron criterion
for magnetism with off-centering second-order Jahn-Teller effect,
all the way to strongly coupled giant magneto-resistance effects
(includes non-oxidic ferrimagnetic semi-conductor thiospinels $Fe
Cr_2 S_4$ and $Fe_{0.5} Cu_{0.5} Cr_2 S_4$, that are $Fe^{2+}$ and
$Fe^{3+}$ end members of solid solution $Fe_{1-x}C_{x}Cr_{2}S_{4}
(0<, =x<,=0.5)$ (Palmer and Greaves 1999)). While lattice
distortions with lowered symmetry reduce competing interactions
(Chern et al 2006; see also Fritsch et al 2004), an insight into the
loss of inversion symmetry comes via the spin-orbit coupling
mechanism which gives the electric polarization P ( $\sim {\mathbf
e} \times {\bf Q}$ ) , where ${\mathbf e}$ is the spin rotation axis
and ${\bf Q}$ is the wave vector of a spiral ) induced upon
transition to a spiral spin-density-wave state triggered by magnetic
frustrations (Mostovoy 2006). Apart from the spin-orbit coupling
factor, a reduction of crystal symmetry (Fd3m to non-centrosymmetric
$F {\bar 4}3m$) in several spinel compounds, including $FeCr_2S_4$
was attributed to a displacement of cations (Mertinat et al 2005,
Charnock et al 1990). Similar off-centering was also found in oxide
spinels (Charnock et al 1990), e.g. magnetite $Fe_3 O_4$.
Additionally, a combination of site-centred (extra holes or
electrons on metal sublattice, e.g. $Fe^{2+}$ and $Fe^{3+}$, where
anions don't play a role) and bond-centred (the alternation of short
and long bonds, in otherwise equivalent sites, lead to a
bond-centered charge density wave) charge-ordering  was suggested
for explaining the multiferroic behavior of $Fe_3 O_4$ below the
Verwey transition at 120K (Khomskii 2004). The co-operative
co-existence of ferroelectric and ferro-magnetic properties in these
structural relatives of greigite --due to a subtle interplay between
charge, spin, orbital and lattice degrees of freedom (Hemberger et
al 2006) --raise the possibility of a similar profile for $Fe_3 S_4$
or close relatives found in enzymes, e.g. $Fe_5 Ni S_8$, for which
no direct evidence is so far available.

% 5.7
\subsection{Preliminary experimental requirements}
What is needed first is a robust model system to explore magnetic
structure formation together with protocols for monitoring
accompanying chemical reactions. Then, the presence of magnetic
rocks in the mound, represented by a surface magnetic field strength
(say, in the range 0-200mT) needs to be checked for any magnetic
structure formation in different concentrations of newly forming
greigite suspension. Here the dispersity of newly forming greigite
clusters whose size range would be expected to closely resemble that
of the FeS dispersion (Fe$_{2}$S$_{2}$ to Fe$_{150}$S$_{150}$) (see
Sect. 5.2) (Rickard and Morse 2005). It could be a reasonable
approximation to mimic the build-up, \textit{for fast-forwarding
geo-time}, by starting out with known (polydisperse) size ranges,
taking into account their initial magnetic susceptibility (along the
lines of Wang and Holm 2003). Further, the 'team-up' of FeS clusters
with organics (see also Rickard and Luther 2007), may well have
deeper roots, as organics play important roles in separate aspects
related to proposed magnetic assemblies, viz., 1) stabilize
colloidal membranes (Russell et al 1994); 2) facilitate
particulation mechanism leading to fractal framboid formation
(Sawlowicz 1993, 2000); 3) enable transformation to greigite in
aqueous dispersed FeS, at the cost of pyrite formation (Rickard et
al 2001) ; and 4) enable generation of metastable phases
intermediate between $Fe Ni_2 S_4$ and $Fe_3 S_4$ (similar to
biological clusters), under mild hydrothermal mound-like conditions
(Tenailleau et al 2006). Thus the inclusion/exclusion of organics
does need to be closely studied in experimental simulations.

% 6.0
\section{Conclusions}
The adaptive nature of biological systems and their fractal
organization cry for a coherent connection between their micro- and
macroscopic domains. A physical basis-the quantum mechanical spin--
for linking the quantum-classical realms at the very origins of life
is suggested in this rudimentary study, rooted in the findings of a
spectrum of scientists (see bibliography). This in turn also helps
to expand the potential of crystal-based theories, and shows how
Life-like dynamics could have been brought about by the magnetic
'face' of minerals. We propose that structured phases with a
magnetic basis for information-transfer, not too far from the mound
(Sect. 5.4), accumulated 'metabolites' (mound-synthesized) riding in
on diffusing super-paramagnetic greigite particles. The evolution of
complexity (biological soft matter with internal degrees of freedom,
asymmetry, organization, etc.) where chemistry was trained to
replace magnetic effects, plus installation/maintenance of energy
transduction mechanisms via energy-rich molecules for using
non-equilibrium sources elsewhere, could have led to the release of
the Ancestor from its H-field providing location. Now, as 'Necessity
is the mother of invention' could it be that the `necessity' for
independence from an increasingly hostile location brought on the
creation of such innovative mechanisms? This possibility seems
intriguing in the light of Patel's findings, where quantum searches
seem to be responsible for the creation of biological language
itself. Moreover, Russell et al have argued that life's hatchery
could have been busy by 3.8 Gyr, evolving fast enough for a branch
to have reached the ocean surfaces by 3.5 Gyr, as evidenced by
photosynthetic signatures. The gestation period of life had to have
been less than the umbilical mound's delivery of the formative
hydrothermal solution, i.e., certainly less than 3 million years,
and probably less than 30,000 years (Fr\"uh-Green et al. 2003).
Indeed, a magnetic start to Life could provide the ingredients for
an intelligent Ancestor, along the lines envisaged by Lloyd (2006)
for a computing universe. Again, it seems to be a physically
feasible embodiment (Mitra-Delmotte and Mitra 2007; 2009) of Paul
Davies's Q-Life proposal (2008), as also acknowledged by him in
Merali (2007). A magnetic basis of assembly could also offer
robustness to an `open' system against interference from a
decohering environment. On the other hand, as evidence of quantum
processing effects in biology trickles in, it appears that Nature is
equipped for tackling environmental intrusion. Sure enough, with
regard to Brownian noise, Nature seems to know how to not only
overcome adversity, but instead put it to its advantage by
harnessing it. At the other-macroscopic-end too, elegant examples
can be seen in the seed dispersal strategies that use this very
'intrusion' by the environment (wind, water, or even creatures).
Thus, the environment apparently provides feedback to the adaptive
living system, besides defining 'necessity' and acting as a
'watch-dog' leading to new nodes in biological evolution (McFadden
and Al-Khalili 1999) (Sect. 2.1). Could it be that the paradigm of
environment-decoherence being a big obstacle against quantum
processing events in biology, needs to be reviewed since
environmental interference itself seems to be an active component of
Nature's search technique?

{\bf Acknowledgements}: One of us (ANM) is grateful to Prof
Krishnaswami Alladi for this opportunity to be associated with this
memorial volume dedicated to (the late) Professor Alladi
Ramakrishnan. The theme of the article has been governed by a desire
to conform to his versatile interest in an entire gamut of physical
science through an appropriate choice of subject. The latter comes
from a recent father-daughter (nay daughter-father!) collaboration
seeking a $Magnetic$ Origin of Life, a subject which represents an
ultimate synthesis of physics with biological chemistry through the
complex terrain of geological science. We thank Prof.M.J.Russell for
inspiration and constant support (data and key references); Prof.Z.
Sawlowicz for key references; Dr. A.Boyce for active help with his
labelled framboid-in-chimney pictures; Prof. K. Matsuno for
suggesting a closer look at electrostatic effects; Prof. A.K. Pati
for bringing "Quantum Aspects of Life" to our notice. This work was
entirely financed, with full infrastructural support, by Dr.
Jean-Jacques Delmotte; Drs A. Bachhawat and B.Sodermark gave a
gentle push; Dr. V. Ghildyal and Mr. Vijay Kumar helped with
manuscript processing.

\end{document}